\documentclass[12pt,twosde]{article}  
\usepackage{cmp2e,amssymb,psfig,epsfig} 

\title{Interacting $N$-vector order parameters with $O(N)$ symmetry}
\author{Andrea Pelissetto\refaddr{label1},
       Ettore Vicari\refaddr{label2}} 

\addresses{
\addr{label1} 
Dip. Fisica dell'Universit\`a di Roma ``La Sapienza"  and INFN, 
P.le Moro 2, I-00185 Roma, Italy
\addr{label2} 
Dip. Fisica dell'Universit\`a di Pisa
and INFN, V. Buonarroti 2, I-56127 Pisa, Italy
} 

\begin{document}

\maketitle

\begin{abstract}
We consider the critical behavior of the most general system of two $N$-vector 
order parameters that is $O(N)$ invariant.
We show that it may a have a multicritical transition with 
enlarged symmetry controlled by the chiral $O(2)\otimes O(N)$ fixed point. 
For $N=2$, 3, 4, if the system is also invariant under the exchange of the 
two order parameters and under independent parity transformations, one 
may observe a critical transition controlled by a fixed point belonging 
to the $mn$ model. Also in this case there is a symmetry enlargement at the 
transition, the symmetry being $[SO(N)\oplus SO(N)]\otimes C_2$, where $C_2$ 
is the symmetry group of the square.

\pacs 05.70.Jk, 64.60.Fr, 75.10.Hk.
\end{abstract}


\section{Introduction} \label{sec1}

The critical behavior of a system with a single $N$-vector order
parameter is well known \cite{ZJ-libro,PV-rev}.  In this paper we
investigate the critical behavior of a system with two $N$-vector
parameters that is invariant under $O(N)$ transformations and
independent parity transformations.

If the two $N$-vector order parameters are identical, i.e. the model is 
symmetric under their exchange, the most general 
Landau-Ginzburg-Wilson (LGW) $\Phi^4$ Hamiltonian is given by
\begin{eqnarray}
{\cal H}_{\rm cr} &=& \int d^3 x \left[
  {1\over2}
\sum_\mu (\partial_\mu \phi \cdot \partial_\mu\phi + 
          \partial_\mu \psi \cdot \partial_\mu\psi  ) + 
  {r\over2} (\phi^2 + \psi^2) + \right.
\nonumber \\
 && \left. + {u_0\over4!} (\phi^4 + \psi^4) + 
      {w_0\over4!} \phi^2 \psi^2 + 
      {z_0\over4!} (\phi\cdot\psi)^2 \right],
\label{Hcr}
\end{eqnarray}
where $\psi_a$ and $\phi_a$ are two $N$-dimensional vectors. This
Hamiltonian is well defined for $u_0>0$, $2 u_0 +
w_0 > 0$, and $2 u_0 + w_0 + z_0 > 0$.  
Hamiltonian (\ref{Hcr}) is invariant under the transformations
\begin{equation}
  ({\mathbb Z}_2)_{\rm exch} \otimes 
  ({\mathbb Z}_2)_{\rm par} \otimes O(N).
\end{equation}
The first ${\mathbb Z}_2$ group is related to the exchange transformations
$\phi\leftrightarrow\psi$, while the second group is related to the 
parity transformations $\phi \to -\phi$, $\psi \to \psi$, or, equivalently,
$\phi\to\phi$, $\psi\to -\psi$ (note that the transformation 
$\phi\to-\phi$, $\psi\to-\psi$ is already accounted for by the $O(N)$ 
group). 

If the two order parameters are not identical and therefore
the symmetry is only
\begin{equation}
({\mathbb Z}_2)_{\rm par} \otimes O(N), 
\end{equation}
the corresponding LGW $\Phi^4$ Hamiltonian is 
\begin{eqnarray}
{\cal H}_{\rm mcr} &=& \int d^3 x \left[
  {1\over2}
\sum_\mu (\partial_\mu \phi \cdot \partial_\mu\phi + 
          \partial_\mu \psi \cdot \partial_\mu\psi  ) + 
  {r_1\over2} \phi^2 + {r_2\over2} \psi^2 + \right.
\nonumber \\
 && \left. + {u_0\over4!} \phi^4 + {v_0\over4!} \psi^4 + 
      {w_0\over4!} \phi^2 \psi^2 + 
      {z_0\over4!} (\phi\cdot\psi)^2 \right],
\label{Hmcr}
\end{eqnarray}
that is well defined for $u_0 > 0$, $v_0 > 0$, $w_0 + 2 \sqrt{u_0 v_0}
> 0$, and $w_0 + z_0 + 2 \sqrt{u_0 v_0} > 0$. Hamiltonian (\ref{Hmcr})
has two different mass terms and thus it gives rise to a variety of
critical and multicritical behaviors.  It generalizes the
multicritical Hamiltonian considered in Ref.~\cite{KNF-76} that has
$w_0 = 0$ and is symmetric under the larger symmetry group $O(N)\oplus
O(N)$.  

For $N=2$ there is a transformation of the fields and couplings 
that leaves invariant Hamiltonians (\ref{Hcr}) or (\ref{Hmcr}).
If we transform the fields as $\phi'_a =
\sum_b\epsilon_{ab} \phi_b$, $\psi'_a = \psi_a$ and the couplings as
\begin{equation}
u'_0 = u_0, \qquad v'_0 = v_0, \qquad w'_0 = w_0 + z_0, \qquad z'_0 = - z_0, 
\label{symmetry}
\end{equation}
we reobtain Hamiltonians (\ref{Hcr}) and (\ref{Hmcr}) expressed in 
terms of the primed fields and couplings \cite{CPV-04}. 
This implies that, for any FP with $z > 0$, there exist an equivalent 
one with the same stability properties and $z < 0$.

Finally, if we do not require the invariance of the model under independent
parity transformations, i.e., the model is only 
$O(N)$ symmetric, we must add an additional quadratic term
$\phi\cdot \psi$ and two additional quartic terms, $(\phi\cdot \psi)
\phi^2$ and $(\phi\cdot \psi) \psi^2$. In this case, the general
analysis becomes more complex since we have to deal with a
multicritical theory with three quadratic terms.

In this paper we investigate whether
theories (\ref{Hcr}) and (\ref{Hmcr}) have stable fixed points (FPs)
in three dimensions.
We do not determine the renormalization-group (RG) flow in the full theory,
but rather we show that stable FPs can be identified by an 
analysis of the submodels whose RG flow is already known.
We consider the stable FPs of the submodels and 
determine their stability properties with respect to
the perturbations that are present in the complete theory.
In this way we are able to identify three stable FPs. For any $N$, 
there is an $O(2)\otimes O(N)$ symmetric FP. This FP may be the 
relevant one for models with $z_0 > 0$,
which may therefore show 
a symmetry enlargement at the (multi)critical transition.
For $N=2$ there is an equivalent $O(2)\otimes O(2)$ symmetric FP with $z < 0$,
a consequence of symmetry (\ref{symmetry}), which 
may be the relevant one for models with $z_0 < 0$. 
For $N=2,3,4$ we find that ${\cal H}_{\rm cr}$---but not the 
multicritical theory ${\cal H}_{\rm mcr}$--- has 
another stable FP that belongs to the 
so-called $mn$ model \cite{Aharony-76} with $n = 2$ and $m=N$.
Also in this case there is a symmetry enlargement at the transition:
the FP is symmetric under the group $[SO(N)\oplus SO(N)]\otimes C_2$
where $C_2$ is the symmetry group of the square.
It is interesting to note that the chiral $O(2)\otimes O(N)$ FP is also 
stable if we do not require the model to be invariant under independent 
parity transformations. Indeed, the additional terms
$(\phi\cdot \psi) \phi^2$ and 
$(\phi\cdot \psi) \psi^2$ are irrelevant perturbations at the chiral FP.

In the analysis we mainly use the
minimal-subtraction ($\overline{\rm MS}$) scheme without $\epsilon$
expansion (henceforth indicated as $3d$-$\overline{\rm MS}$
scheme) in which no $\epsilon$ expansion is performed and $\epsilon$
is set to the physical value $\epsilon=1$ \cite{SD-89}. In order 
to generate the relevant perturbative series we use
a symbolic manipulation program that generates the diagrams and 
computes symmetry and group factors. For the Feynman integrals
we use the results reported in Ref.~\cite{KS-01}.
In this way we obtained five-loop $3d$-${\overline{\rm MS}}$ expansions.

\section{Mean-field analysis} \label{sec2}

\begin{figure}[t]
\centerline{\epsfig{file=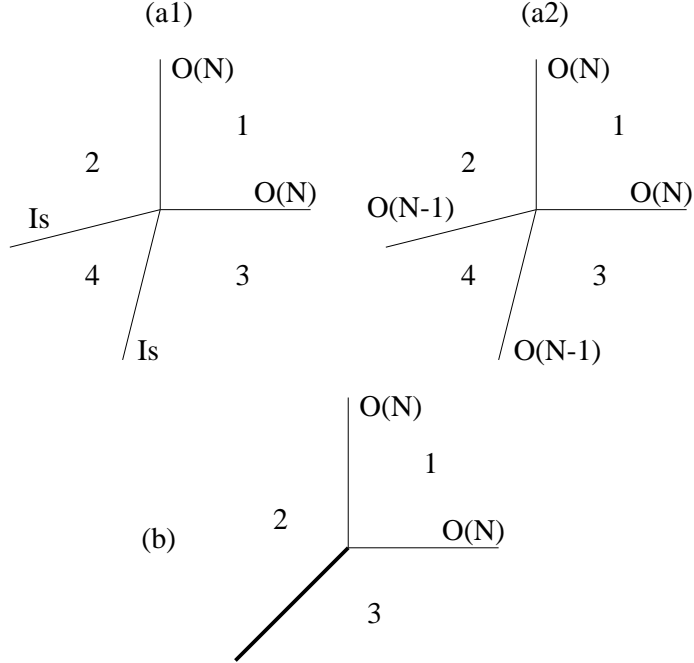,width=10truecm}}
\caption{Possible multicritical phase diagrams.
Thin lines indicate second-order transitions, while the thick line in case (b)
corresponds to a first-order transition.
``Is" indicates an Ising transition.
}
\label{fig-MF}
\end{figure}

The mean-field analysis of the critical behavior of Hamiltonian 
${\cal H}_{\rm cr}$ is quite straightforward. If $r > 0$ 
the system is paramagnetic, with $\phi = \psi = 0$. 
For $r < 0$ there are three possible low-temperature phases:
\begin{itemize}
\item[(a)] For $w_0 > 2 u_0$ and $z_0 > 2 u_0 - w_0$, we have 
$\phi\not=0$ and $\psi = 0$ (or viceversa). The corresponding 
symmetry-breaking pattern is 
$ ({\mathbb Z}_2)_{\rm exch} \otimes 
  ({\mathbb Z}_2)_{\rm par} \otimes O(N) \to
  ({\mathbb Z}_2)_{\rm par} \otimes O(N-1)$. 
\item[(b)] For $z_0 < 0$ and $-2 u_0 < w_0 + z_0 < 2 u_0$, we have 
$\phi=\psi\not=0$. The corresponding 
symmetry-breaking pattern is 
$ ({\mathbb Z}_2)_{\rm exch} \otimes 
  ({\mathbb Z}_2)_{\rm par} \otimes O(N) \to
  ({\mathbb Z}_2)_{\rm exch} \otimes O(N-1)$. 
\item[(c)] For $z_0 > 0$ and $-2 u_0 < w_0 < 2 u_0$, we have 
$|\phi|=|\psi|\not=0$, $\phi\cdot\psi=0$. The corresponding 
symmetry-breaking pattern is 
$ ({\mathbb Z}_2)_{\rm exch} \otimes 
  ({\mathbb Z}_2)_{\rm par} \otimes O(N) \to
  ({\mathbb Z}_2)_{\rm exch} \otimes O(N-1)$. 
\end{itemize}
The analysis of the mean-field behavior of ${\cal H}_{\rm mcr}$ is presented 
for $N=4$ in Ref.~\cite{BPV-03} and it is easily extended to the present
case. There are three possible phase diagrams:

\begin{itemize}
\item[(a1)] For $z_0 < 0$ and 
$- 2 \sqrt{u_0 v_0} < w_0 + z_0 < 2 \sqrt{u_0 v_0}$,
the multicritical point is tetracritical, see Fig.~\ref{fig-MF}.
Phase 1 is paramagnetic with $\phi=\psi=0$, 
in phase 2 $\phi\not=0$ and $\psi=0$, while in phase 3 
the opposite holds, $\phi =0$ and $\psi\not=0$;
in phase 4 $\phi\not=0$, $\psi\not=0$ with $\phi\| \psi$.
All transitions are of second order. 
Transitions 1-2 and 1-3 are
associated with the symmetry breaking 
${\mathbb Z}_2 \otimes O(N)\to {\mathbb Z}_2 \otimes O(N-1)$. 
In the presence of fluctuations these transitions belong to the 
$O(N)$ universality class.  Transitions 2-4 and 3-4 are 
associated with the symmetry-breaking pattern
${\mathbb Z}_2 \otimes O(N-1)\to 
O(N-1)$. In the presence of fluctuations
they should belong to the Ising universality class.

\item[(a2)] For $z_0 > 0$ and $- 2 \sqrt{u_0 v_0} < w_0 < 2 \sqrt{u_0 v_0}$ 
the multicritical point is tetracritical, see Fig.~\ref{fig-MF}.
Phases 1, 2, and 3 as well as transitions 
1-2 and 1-3 are identical to those discussed in case (a1).
In phase 4 $\phi\not=0$, $\psi\not=0$ with $\phi\cdot\psi=0$.
All transitions are second-order ones. Transitions 2-4 and 3-4 are 
associated with the symmetry-breaking pattern
${\mathbb Z}_2 \otimes O(N-1)\to 
{\mathbb Z}_2 \otimes O(N-2)$ and, in the presence of fluctuations,
they should belong to the $O(N-1)$ universality class.

\item[(b)] For $w_0 > 2 \sqrt{u_0 v_0}$ and $w_0 + z_0 > 2 \sqrt{u_0 v_0}$ 
the multicritical point is bicritical, see Fig.~\ref{fig-MF}.
Phases 1, 2, and 3 as well as transitions 
1-2 and 1-3 are identical to those discussed in case (a1).
The transition between phases 2 and 3 is of first order.

\end{itemize}

\section{Analysis of some particular cases}
\label{sec3}

\subsection{Particular models and fixed points}
\label{sec3.1}

The three-dimensional properties of the RG flow are determined by its
FPs.  Some of them can be identified by considering particular cases
in which some of the quartic parameters vanish.  For 
${\cal H}_{\rm cr}$ we can easily recognize two 
submodels:
\begin{itemize}
\item[(a)] 
The $O(2)\otimes O(N)$ model with Hamiltonian \cite{Kawamura-88}
\begin{eqnarray}
{\cal H}_{\rm ch} =&& \int d^d x \left\{ 
  {1\over2}
\sum_{ai} \left[ \sum_\mu (\partial_\mu \Phi_{ai})^2 + r \Phi_{ai}^2 
      \right]   
\right.
\nonumber \\
&& \left.
+ {g_{1,0}\over 4!}  ( \sum_{ai} \Phi_{ai}^2)^2 
+ {g_{2,0}\over 4!}  \left[ \sum_{i,j} 
( \sum_a \Phi_{ai} \Phi_{aj})^2 - (\sum_{ai} \Phi_{ai}^2)^2 \right]
\right\} ,
\label{Hch}
\end{eqnarray}
where $\Phi_{ai}$ is an $N\times2$ matrix, i.e., $a=1,\ldots,N$ and $i=1,2$.
Hamiltonian (\ref{Hcr}) reduces to (\ref{Hch}) for $2 u_0 - w_0 - z_0 = 0$,
if we set $\Phi_{a1}=\phi_a$, $\Phi_{a2} = \psi_a$, 
$u_0 = g_{1,0}$, $w_0 = 2 (g_{1,0} - g_{2,0})$, and $z_0 = 2 g_{2,0}$. 
The properties of $O(2)\otimes O(N)$ models are reviewed in
Refs.~\cite{Kawamura-98,PV-rev,DMT-03,CPPV-04}.  In three
dimensions perturbative calculations within the three-dimensional
massive zero-momentum (MZM)
scheme \cite{PRV-01,CPS-02} and within the $3d$-$\overline{\rm
MS}$ scheme \cite{CPPV-04} indicate the presence of a stable FP with
attraction domain in the region $g_{2,0}>0$ for all values of $N$
(only for $N=6$ the evidence is less clear: a FP is identified in 
the $3d$-$\overline{\rm MS}$ scheme but not in the MZM scheme).
For $N=2$, these conclusions have been
recently confirmed by a Monte Carlo calculation \cite{CPPV-04}.  On
the other hand, near four dimensions, a stable FP is found only for
large values of $N$, i.e., $N> N_c= 21.80 - 23.43 \epsilon + 7.09
\epsilon^2 + O(\epsilon^3)$ \cite{Kawamura-88,ASV-95,PRV-01b,CP-03}.
A stable FP with attraction domain in the region $g_{2,0}<0$ exists
for $N=2$ (it belongs to the XY universality class)
\cite{Kawamura-88}, for $N=3$ (Ref.~\cite{DPV-04}), and
$N=4$ (Ref.~\cite{CPV-04}).  Note that nonperturbative approximate RG
calculations have so far found no evidence of stable FPs for $N=2$ and
3 \cite{TDM-00,DMT-03,KW-01}. In the following we call the FP with $g_2>
0$ {\em chiral} FP (we indicate it with 
$g_{1,\rm ch}^*$, $g_{2,\rm ch}^* > 0$), 
while the FP with $g_2 < 0$ is named {\em
collinear} FP and indicated with 
$g_{1,\rm cl}^*$, $g_{2,\rm cl}^* < 0$. 

\item[(b)] 
The so-called $mn$ model with Hamiltonian \cite{Aharony-76}
\begin{eqnarray}
{\cal H}_{\rm mn} =&& \int d^d x \left\{ 
  {1\over2}
\sum_{ai} \left[ \sum_\mu (\partial_\mu \Phi_{ai})^2 + r \Phi_{ai}^2 
      \right]   
\right.
\nonumber \\
&& \qquad\qquad \left.
+ {g_{1,0}\over 4!}  ( \sum_{ai} \Phi_{ai}^2)^2 
+ {g_{2,0}\over 4!}   \sum_{abi} 
 \Phi_{ai}^2 \Phi^2_{bi}
\right\} ,
\label{HMN}
\end{eqnarray}
where $\Phi_{ai}$ is an $m\times n$ matrix, i.e., $a=1,\ldots,m$ and 
$i=1,\ldots,n$.
Hamiltonian (\ref{Hcr}) reduces to (\ref{Hch}) for $n=2$, $m=N$, and 
$z_0 = 0$, if we set $\Phi_{a1}=\phi_a$, $\Phi_{a2} = \psi_a$, 
$u_0 = g_{1,0} + g_{2,0}$ and $w_0 = 2 g_{1,0}$.
A stable FP is the $O(m)$ FP
with $g_1 = 0$ and $g_2 = g_m^*$,
where $g_m^*$ is the FP value of the renormalized coupling
in the $O(m)$ model. In App.~\ref{AppA} we show that the model
has a second stable FP with $g_2 < 0$ for $n = 2$ and $m=2$, 3, and 4.
We name this FP the $mn$ FP and we label the corresponding coordinates by
$g_{1,mn}^*$ and $g_{2,mn}^*$.
\end{itemize}
The presence of these two submodels that have one parameter less than
the original model imply that the quartic parameter space 
splits into four regions such that the RG flow does not cross the 
two planes $z = 0$ and $2 u - w - z = 0$. Note that, for $N=2$, because
of symmetry (\ref{symmetry}), we should only consider the region $z\ge 0$.

The results for models (a) and (b) allow us
to identify four possible FPs that are candidates for being stable FPs
of the full theory:
\begin{itemize}
\item[(1)] $u = g^*_{1,\rm ch}$, 
           $w = 2 (g^*_{1,\rm ch} - g^*_{2,\rm ch})$, 
           $z = 2 g^*_{2,\rm ch}$; 
this FP may be the stable FP of the trajectories that start in the region
$z_0 > 0$. 
\item[(2)] $u = g^*_{1,\rm cl}$, 
           $w = 2 (g^*_{1,\rm cl} - g^*_{2,\rm cl})$, 
           $z = 2 g^*_{2,\rm cl}$; 
this FP may be the stable FP of the trajectories that start in the region
$z_0 < 0$. 
\item[(3)] $u = g^*_{N}$, $w = 0$, $z = 0$;
this FP may be the stable FP of the trajectories that start in the region
$2 u_0 - w_0 - z_0 > 0$. 
\item[(4)] $u = g^*_{1,mn} + g^*_{2,mn}$, $w =  2 g^*_{1,mn}$, $z = 0$ for $N=2,3,4$;
this FP may be the stable FP of the trajectories that start in the region
$2 u_0 - w_0 - z_0 < 0$.
\end{itemize}
Note that, because of symmetry (\ref{symmetry}), for $N=2$ there is also 
a chiral (resp. collinear) FP with $z < 0$ (resp. $z > 0$). 

The analysis of the particular cases of Hamiltonian (\ref{Hmcr}) is 
very similar. There are two relevant submodels:
\begin{itemize}
\item[(a)]
The forementioned $O(2)\otimes O(N)$ model for 
$u_0 = v_0$ and $w_0 + z_0 = 2 u_0$. The identication is 
$u_0 = v_0 = g_{1,0}$, $w_0 = 2 g_{1,0} - 2 g_{2,0}$, and 
$z_0 = 2 g_{2,0}$. 
\item[(b)]
The $O(N)\oplus O(N)$ model \cite{KNF-76}:
\begin{eqnarray}
{\cal H}_{\rm mcr,2} =&& \int d^3 x \left[
  {1\over2}
\sum_\mu (\partial_\mu \phi \cdot \partial_\mu\phi + 
          \partial_\mu \psi \cdot \partial_\mu\psi  ) + 
  {r_1\over2} \phi^2 + {r_2\over2} \psi^2 + \right.
\nonumber \\
 && \left. + {f_{1,0}\over4!} \phi^4 + {f_{2,0}\over4!} \psi^4 + 
      {f_{3,0}\over4!} \phi^2 \psi^2 \right].
\label{HmcrNN}
\end{eqnarray}
Hamiltonian ${\cal H}_{\rm mcr}$ reduces to this model
for $z_0 = 0$, with the obvious identification of the parameters.
Hamiltonian (\ref{HmcrNN}) describes the multicritical behavior of a model 
with two $N$-vector order parameters that is symmetric under independent
$O(N)$  transformations of the two order parameters, i.e. 
that is invariant under the symmetry group $O(N)\oplus O(N)$
\cite{KNF-76}.  In the case we are interested in, i.e. for $N\geq 2$,
the stable FP is the decoupled FP \cite{CPV-03,Aharony-02}, i.e., 
$f_{3} = 0$, $f_1 = f_2 = g^*_N$ (see also App.~\ref{AppA}).
\end{itemize}
Note that in this case the $O(2)\otimes O(N)$ model has two parameters less 
than the original one and thus its presence does not 
imply any separation of the RG flow. Instead, the second model implies 
that the quartic parameter space 
splits into two regions such that the RG flow does not cross the 
plane $z = 0$. The analysis of the possible FPs is identical 
to that presented above, since the FPs we have identified are exactly 
those we have already described.

\subsection{Stability of the O(2)$\otimes$O($N$) fixed points}
\label{sec3.2}

In this section we study the stability properties of the two FPs that
appear in the $O(2)\otimes O(N)$ model.
For this purpose we need to classify the
perturbations of the $O(2)\otimes O(N)$ model that do not break the 
$O(N)$ invariance.
The multicritical Hamiltonian (\ref{Hmcr}) can be
rewritten as
\begin{equation}
{\cal H}_{\rm mcr} = {\cal H}_{\rm ch} + {1\over2} r_{2,2} V^{(2,2)} + 
      {1\over 4!} f_{4,4} V^{(4,4)} + 
      {1\over 4!} f_{4,2} V^{(4,2)},
\end{equation}
where ${\cal H}_{\rm ch}$ is the $O(2) \otimes O(N)$-symmetric
Hamiltonian (\ref{Hch}) with $r = (r_1 + r_2)/2$, 
$g_{1,0} = (2 u_0 + 2 v_0 + w_0 + z_0)/6$, 
$g_{2,0} = (2 u_0 + 2 v_0 -2 w_0 + z_0)/6$, and
$r_{2,2} = (r_1 - r_2)/2$, $f_{4,2} = (u_0 - v_0)/2$, and 
$f_{4,4} = (u_0 + v_0 - w_0 - z_0)/6$. 
Here, $V^{(2,2)}$, $V^{(4,4)}$, and $V^{(4,2)}$, 
are respectively a quadratic term that transforms as a spin-2 operator
under the $O(2)$ group and two quartic terms that transform as a 
spin-4 and as a spin-2 operator respectively.  Their explicit expressions are:
\begin{eqnarray}
&& V^{(2,2)} \equiv \phi^2 - \psi^2, \nonumber \\
&& V^{(4,2)} \equiv (\phi^2 + \psi^2) V^{(2,2)}, \nonumber \\
&& V^{(4,4)} \equiv (\phi^2)^2 + (\psi^2)^2 - 
         2 \phi^2 \psi^2 - 4 (\phi\cdot\psi)^2.
\end{eqnarray}
A detailed description of all possible perturbations of the 
$O(2)\otimes O(N)$ FP the leave invariant the $O(N)$ group
can be found in App.~B of Ref.~\cite{CPV-04}.  
Note that for ${\cal H}_{\rm cr}$ we have $r_{2,2} = f_{4,2} = 0$, so that 
one must only consider the spin-4 quartic perturbation.

Let us first discuss the chiral FP (a) that has $g^*_{2,\rm ch} > 0$.  
In order to estimate the RG dimensions $y_{4,2}$ and
$y_{4,4}$ of the above-reported perturbations, we computed the corresponding
five-loop $\overline{\rm MS}$ series and we analyzed
them within the $3d$-$\overline{\rm MS}$ scheme.  

\begin{table}
\begin{center}
\begin{tabular}{|llll|}
\hline\hline
\multicolumn{1}{|l}{}& 
\multicolumn{1}{c}{$y_{2,2}$}&
\multicolumn{1}{c}{$y_{4,2}$}&
\multicolumn{1}{c|}{$y_{4,4}$} \\
\hline
ch,2 & 1.34(15) & $-$1.6(1.1) &  $-$0.9(4)    \\
ch,3 & 1.21(9)  & $-$1.4(8)   &  $-$1.0(3)    \\
ch,4 & 1.17(8)  & $-$1.1(5)   &  $-$1.0(3)    \\
ch,6 & 1.13(9)  & $-$0.9(4)   &  $-$0.9(2) \\
ch,8 & 1.13(8)  & $-$0.9(5)   &  $-$0.9(2)   \\
ch,16 & 1.08(2) & $-$1.0(3)  &  $-$0.94(11) \\
ch,$\infty$ & 1 &  $-1$  & $-1$ \\
cl,2 & 1.9620(8) & \hphantom{$-$}0   & 0.532(12) \\
cl,3 & 2.05(15)  &    & 0.9(3) \\
cl,4 & 2.05(15)  &    & 0.9(7) \\
\hline
\end{tabular}
\end{center}
\caption{ 
Estimates of the RG dimensions $y_{2,2}$, $y_{4,2}$, and
$y_{4,4}$ of the operators $V^{(2,2)}$, $V^{(4,2)}$, and
$V^{(4,4)}$ at the chiral (ch) FP and at the collinear (cl) FP.
They have been obtained from a conformal-mapping analysis of the 
corresponding $3d$-$\overline{\rm MS}$ 
5-loop perturbative expansions.
The results for $y_{2,2}$ are taken from 
Ref.~\protect\cite{CPV-04}. The results at the collinear FP for 
$N=2$ have been computed by using the mapping with the XY model and the 
results of Ref.~\cite{CPV-03}.
}
\label{tabn2}
\end{table}

The perturbative series, that are not reported here but are available
on request, were analyzed using the conformal-mapping method and the
Pad\'e-Borel method, following closely Ref.~\cite{CPV-00}, to
which we refer for details.  The error on the conformal-method results
takes into account the spread of the results as the parameters
$\alpha$ and $b$ are varied (cf.~Ref.~\cite{CPV-00} for definitions)
and the error due to the uncertainty of the FP location (we use the
estimates reported in Refs.~\protect\cite{PRV-01,CPS-02,CPPV-04,CPV-04}). 
The results of the analyses using the conformal-mapping method 
are reported in Table~\ref{tabn2}.  Completely consistent results are 
obtained by using Pad\'e-Borel approximants. As it can be seen, 
$y_{4,2}$ and $y_{4,4}$ are always negative, indicating that the 
chiral FP is stable for any $N$. This FP is 
therefore expected to be the relevant FP whenever the RG flow starts in
the region $z_0 > 0$. For $N=2$, symmetry (\ref{symmetry}) implies that 
a chiral FP [the equivalent one that is obtained by using (\ref{symmetry})] 
may also be reached from the region $z_0 < 0$. 

It is of interest to compute also the RG dimension $y_{2,2}$ 
of the quadratic perturbation. 
For the multicritical Hamiltonian it is related to the crossover exponent 
$\phi$: $\phi=\nu y_{2,2}$, where $\nu$ is the correlation-length
exponent at the chiral FP (see Refs.~\cite{PRV-01,CPS-02,CPPV-04} for numerical
estimates). The exponent $y_{2,2}$ has already been computed for several 
values of $N$ in Ref.~\cite{CPV-04}: indeed, $y_{2,2} = y_4$, where 
$y_4$ is the RG dimension of the operator $O^{(4)}$ defined in 
App.~C of Ref.~\cite{CPV-04}. Numerical estimates, taken from 
Ref.~\cite{CPV-04}, are reported in Table~\ref{tabn2}.

Now, let us consider the collinear FP (b) that has $g_{2,\rm cl}^*(N)
< 0$ for $2 \le N \le 4$. 
For $N=2$ the RG dimensions at the collinear FP can be related
to the RG dimensions of operators in the XY model. Indeed,
the O(2)$\otimes$O(2) collinear FP is equivalent to an XY FP.
The mapping is the following.
One defines two fields $a_i$ and $b_i$, $i=1,2$, and considers
\cite{Kawamura-88}
\begin{eqnarray}
&& \phi_{11} = (a_1 - b_2)/\sqrt{2},
\nonumber \\
&& \phi_{22} = (a_1 + b_2)/\sqrt{2},
\nonumber \\
&& \phi_{12} = (b_1 - a_2)/\sqrt{2},
\nonumber \\
&& \phi_{21} = (b_1 + a_2)/\sqrt{2}.
\end{eqnarray}
At the collinear FP, fields $a$ and $b$ represent two independent XY fields.
Using this mapping it is easy to show that
$V^{(4,2)} \sim {\cal O}^{(3,1)}_i(a) b_j$, ${\cal O}^{(3,1)}_i(b) a_j$, and 
$V^{(4,4)} \sim T_{11}(a) T_{11}(b)$, $T_{12}(a) T_{12}(b)$, where
\begin{equation}
{\cal O}^{(3,1)}_i(a) \equiv a_i a^2, \qquad\qquad 
T_{ij}(a) \equiv  a_i a_j - {1\over2} \delta_{ij} a^2 .
\end{equation}
Thus, if $y_{3,1}$ and $y_2$ are the RG dimensions of 
${\cal O}^{(3,1)}_i$ and $T_{ij}$ in the XY model, we have 
\begin{equation}
y_{4,2} = y_h + y_{3,1} - 3, \qquad\qquad 
y_{4,4} = 2 y_2 - 3.
\end{equation}
By using the equations of motion, one can relate ${\cal O}^{(3,1)}_i$ to 
$a_i$ \cite{DPV-03} and obtain $y_{3,1} = 3 - y_h$, so that 
$y_{4,2} = 0$ exactly (this holds in three dimensions; in generic 
dimension $d$, $y_{4,2} = 3-d$). For $y_2$ we can use the result reported in 
Ref.~\cite{CPV-03}, obtaining $y_{4,4} = 0.532(12)$. The analysis
of the perturbative series gives results that are fully consistent:
$y_{4,2} = 0.0(1)$, $y_{4,4} = 0.57(4)$. 

In order to determine $y_{4,2}$ and $y_{4,4}$ for $N=3$ and 4 we  
analyzed the corresponding 5-loop $3d$-$\overline{\rm MS}$ expansions.
The results for $y_{4,4}$ are reported in Table~\ref{tabn2}. 
They indicate that $y_{4,4}$ is positive, which implies that the spin-4 quartic 
perturbation is relevant and therefore the collinear FP is unstable. 
We do not quote any result for $y_{4,2}$. The perturbative analysis 
does not allow us to obtain any reliable result: the estimates 
vary significantly with the parameters $b$ and $\alpha$ and with the 
perturbative order. 

It is interesting to observe that, on the basis of the group-theoretical 
analysis reported in App. B of Ref.~\cite{CPV-04}, any quartic 
perturbation of the 
chiral FP that leaves invariant the $O(N)$ symmetry is a combination
of spin-2 and spin-4 operators. Thus, the results presented here 
indicate that the chiral $O(2)\otimes O(N)$ FP is stable under 
{\em any} perturbation that preserves the $O(N)$ symmetry. 
In particular, it is also stable
under a perturbation of the form $(\phi\cdot\psi)(a \phi^2 + b\psi^2)$
that may arise if the model is not invariant under independent parity
transformations. Indeed, such a term is nothing but a particular 
combination of spin-2 and spin-4 perturbations. In the 
notations of App.~B of Ref.~\cite{CPV-04} (note that 
$M$ and $N$ of Ref.~\cite{CPV-04} should be replaced by $N$ and 2 
respectively) we have
\begin{equation} 
(\phi\cdot\psi)(a \phi^2 + b\psi^2) = 
    {1\over2} (a+b) {\cal O}^{(4,2,1)}_{12} + 
    {1\over3} (a-b) {\cal O}^{(4,4)}_{1112} .  
\end{equation}
Moreover, the additional quadratic term $\phi\cdot\psi$ is nothing
but a component of the spin-2 quadratic term that breaks the $O(2)$
group, so that 
the associated crossover exponent is again $\phi = \nu y_{2,2}$,
$y_{2,2}$ being reported in Table~\ref{tabn2}.
This is a general result that follows from the analysis of 
Ref.~\cite{CPV-04}: any quadratic perturbation of the 
$O(2)\otimes O(N)$ FP that does not break the $O(N)$ invariance
is a combination of the components of the spin-2 quadratic 
operator. Thus, any perturbation is always associated with
the same crossover exponent $\phi = \nu y_{2,2}$.

\subsection{Stability of the decoupled $O(N)\oplus O(N)$
fixed point}
\label{sec3.3}

We now consider FP (3) discussed in
Sec.~\ref{sec3.1}.  In order to check its stability, we
must determine the RG dimensions at the FP of the
perturbations
\begin{equation}
P_E \equiv \phi^2 \psi^2, \qquad 
P_T \equiv  \sum_{ij} T_{\phi,ij} T_{\psi,ij},
\end{equation}
where 
$T_{\phi,ij} = \phi_i \phi_j - {1\over2} \delta_{ij} \phi^2$.
Simple RG arguments show that the RG dimensions 
are given by
\begin{equation}
y_E = {2\over \nu_N} - 3 = {\alpha_N \over \nu_N} ,
\qquad
y_T = {2 y_{2}} - 3,
\label{yet}
\end{equation}
where $\alpha_N$ and $\nu_N$ are the critical exponents of the
3-dimensional $O(N)$ universality class 
(see Ref.~\cite{PV-rev} for a comprehensive review of results), while 
$y_2$ is the exponent associated with the quadratic spin-2 
perturbation in the $O(N)$ model \cite{CPV-02,CPV-03,CPV-O4nova}.

Since $\alpha_N < 0$ for $N\ge 2$ we have $y_E < 0$, i.e. the perturbation 
$P_E$ is always irrelevant. As for $y_T$ we can use the results reported in 
Refs.~\cite{CPV-02,CPV-03,CPV-O4nova}. The spin-2 exponent is equal to 
$y_2 = 1.766(6)$, 1.790(3), 1.813(6) for $N=2$, 3, 4 and increases towards
2 as $N\to\infty$. 
Correspondingly $y_T = 0.532(12)$, 0.580(6), 0.626(12), 
increasing towards 1 as $N\to\infty$.
It follows that $P_T$ is always relevant.
Thus, the decoupled FP is always irrelevant.

\subsection{Stability of the $mn$ fixed point}  \label{sec3.4}

Here, we wish to consider the stability of the $mn$ FP. For this purpose
we must consider the two perturbations
\begin{equation}
P_1 \equiv  (\phi^2)^2 - (\psi^2)^2, \qquad\qquad
P_2 \equiv  (\phi\cdot\psi)^2 - {1\over N} \phi^2 \psi^2.
\end{equation}
Note that $P_1$ is not symmetric under interchange of $\phi$ and $\psi$
and is therefore not of interest for ${\cal H}_{\rm cr}$. The corresponding
RG dimensions $y_1$ and $y_2$ are computed in App.~\ref{AppA}:
$y_1 = 0.4(3)$, 0.2(2), 0.2(2) for $N=2$, 3, 4;
$y_2 = -0.9(5)$, $-$1.0(8), $-$0.8(5) for the same values of $N$.
They indicate that $P_1$ is relevant and $P_2$ is irrelevant at the 
$mn$ FP. Therefore, the $mn$ FP is a stable FP for ${\cal H}_{\rm cr}$
(only $P_2$ should be considered in this case)
and an unstable one for ${\cal H}_{\rm mcr}$.

\section{Conclusions}

In this paper we have investigated the critical behavior of systems described 
by Hamiltonians (\ref{Hcr}) and (\ref{Hmcr}). We find that 
${\cal H}_{\rm cr}$ has three possible stable FPs:
for any $N$, except possibly $N=6$ (for such a value of 
$N$ the evidence of this FP is less robust \cite{CPPV-04}), there is 
the $O(2)\otimes O(N)$ chiral FP that is relevant for systems with 
$z_0 > 0$;
for $N=2$ there is a stable chiral FP with $z < 0$ [equivalent to the 
previous one by symmetry (\ref{symmetry})], that is relevant for
systems with $z_0 < 0$;
for $N=2$, 3, 4,  there is the $mn$ FP that is relevant for systems 
with $2 u_0 - w_0 - z_0 < 0$. 
In the multicritical theory (\ref{Hmcr})
only the chiral FPs are stable.  Thus, systems with $z_0 > 0$ 
(or, for $N=2$, with $z_0 \not= 0$)
may show a multicritical continuous transition with the larger 
$O(2)\otimes O(N)$ symmetry. 

It is interesting to note that the most general $O(N)$-invariant 
LGW Hamiltonian for two $N$-vector parameters includes other couplings 
beside those present in (\ref{Hcr}) and (\ref{Hmcr}). One should consider
\begin{eqnarray}
{\cal H}_{\rm cr,ext} &=& {\cal H}_{\rm cr} + 
{r_2\over2} \phi\cdot \psi + {a_0\over4!} (\phi\cdot \psi) (\phi^2 + \psi^2), 
\label{Hcrext}
\\
{\cal H}_{\rm mcr,ext} &=& {\cal H}_{\rm mcr} + 
{r_3\over2} \phi\cdot \psi + {1\over 4!} (\phi\cdot \psi) 
(a_1 \phi^2 + a_2 \psi^2), 
\end{eqnarray}
depending whether one wants to preserve the symmetry under the exchange of the 
two fields. As we discussed in Sec.~\ref{sec3.2}, the chiral FP is a stable 
FP also for these two extended models. 

Hamiltonians (\ref{Hmcr}) and (\ref{Hcrext}) have two mass parameters 
and thus symmetry enlargement can be observed only at the multicritical point,
where the singular part of the free energy has the form 
\begin{equation}
  F_{\rm sing} = \mu_t^{2-\alpha} f(\mu_g \mu_t^{-\phi}),
\end{equation}
where $\mu_t$ and $\mu_g$ are two linear scaling fields (linear combinations 
of the temperature and of another relevant parameter), $\alpha$ and $\phi$
are the specific-heat and the crossover exponents at the 
$O(2)\otimes O(N)$ model. Note that the same expression,
with the same $\alpha$ and $\phi$, applies to both models,
apart from nonuniversal normalization constants.
In ${\cal H}_{\rm mcr,ext}$ there are three quadratic parameters and thus 
the chiral multicritical point can be observed only if three relevant
parameters are properly tuned. The singular part of the free energy becomes
\begin{equation}
  F_{\rm sing} = \mu_t^{2-\alpha} f(\mu_{g1} \mu_t^{-\phi}, 
         \mu_{g2} \mu_t^{-\phi}),
\end{equation}
where $\mu_{g1}$ and $\mu_{g2}$ are two linear scaling fields associated with
the same crossover exponent $\phi$.

\appendix 

\section{The $mn$ model: new fixed points} \label{AppA}

In this Appendix we consider the $mn$ model defined by Hamiltonian
(\ref{HMN}), focusing on the case $n=2$ that is of interest for the
present paper. Within the $\epsilon$ expansion one finds four FPs, the
stable one being the $O(m)$ FP with $g_1 = 0$ and $g_2 = g^*_m$, where
$g^*_m$ is the FP value of the renormalized zero-momentum coupling in
the $O(m)$ model, see Refs.~\cite{Aharony-76,PV-rev,DHY-04} and
references therein. For $m=2$ (and $n=2$) the $mn$ model is equivalent
\cite{Kawamura-88} to the $O(2)\otimes O(2)$ model defined by
Hamiltonian (\ref{Hch}).  For this model, the results of
Refs.~\cite{PRV-01,CPPV-04} indicate the presence of a new FP that is
not predicted by the $\epsilon$-expansion analysis.  Because of the
mapping, this implies the presence of a new FP in the $mn$ model with
$g_2 < 0$. In the MZM scheme the results of Ref.~\cite{PRV-01} imply
the presence of a stable FP at $g_1 = 4.4(2)$ and $g_2 = - 4.5(2)$,
where the renormalized couplings $g_1$ and $g_2$ are normalized so
that $g_1 = 3 g_{1,0}/(16 \pi R_{2m} m)$, $g_2 = 3 g_{2,0}/(16 \pi R_m
m)$ at tree level ($m$ is the renormalized zero-momentum mass), where
$R_k = 9/(8 + k)$. In the $3d$-$\overline{\rm MS}$ scheme, by using
the results of Ref.~\cite{CPPV-04}, we obtain $g_1 = 2.25(13)$ and
$g_2 = - 2.31(21)$, where $g_i = g_{i,0} \mu^{-\epsilon}/A_d$ with
$A_d = 2^{d-1} \pi^{d/2} \Gamma(d/2)$.
It is thus of interest to check whether additional FPs are also present for 
other values of $m>2$. As we shall show below we find 
an additional FP for $m=3$ and $m=4$. For $m\ge 5$ no new FP is found.

\begin{table}
\begin{center}
\begin{tabular}{|cclllll|}
\hline\hline
\multicolumn{1}{|c}{$m$}& 
\multicolumn{1}{c}{scheme}&
\multicolumn{1}{c}{$g_1$}&
\multicolumn{1}{c}{$g_2$}&
\multicolumn{1}{c}{$p_{\rm FP}$} & 
\multicolumn{1}{c}{$p_{\rm st}$} & 
\multicolumn{1}{c|}{info}  
\\
\hline
2 & $3d$-$\overline{\rm MS}_{5l}$ & 2.3(2) & $-$2.3(2) &  
24/24 & 19/24 & 6/24, 18/24   \\ 

    & $3d$-$\overline{\rm MS}_{4l}$ & 2.4(2) & $-$2.5(3) &  
13/24 & 2/13 & 13/13, 0/24    \\ 

    & MZM$_{6l}$ & 4.60(8) & $-$4.51(11) &  
24/24 & 24/24 & 0/24, 24/24  \\ 

    & MZM$_{5l}$ & 4.7(3) & $-$4.6(4) &  
24/24 & 20/24 & 4/24, 20/24   \\ 
\hline 
&&&&&&\\[-4mm]

3 & $3d$-$\overline{\rm MS}_{5l}$ & 2.5(2) & $-$2.5(2) &   
23/24 & 22/23 & 18/23, 5/23  \\ 

  & $3d$-$\overline{\rm MS}_{4l}$ & 2.5(3) &  $-$2.6(5) &   
13/24 & 2/13 & 13/13, 0/13   \\ 

    & MZM$_{6l}$ & 5.6(3) & $-$5.2(3) &   
23/24 & 23/23 & 13/23, 9/23  \\ 

    & MZM$_{5l}$ & 5.2(2) & $-$4.8(2) &   
24/24 & 24/24 & 1/24, 23/24  \\  
\hline 
&&&&&&\\[-4mm]

4 & $3d$-$\overline{\rm MS}_{5l}$ & 2.9(4) & $-$2.9(3) &   
19/24 & 18/19 & 19/19, 0/19 \\ 

  & $3d$-$\overline{\rm MS}_{4l}$ & 3.0(3)  &  $-$3.0(4) &   
8/24 &  0/8 & 8/8, 0/8
           \\ 

  & MZM$_{6l}$ & 6.6(6) & $-$6.0(6) &   
15/24 & 15/15 & 13/15, 2/15  \\ 

    & MZM$_{5l}$ & 5.9(3) & $-$5.2(3) &   
24/24 & 24/24 & 12/24, 12/24 \\ 
\hline
\end{tabular}
\end{center}
\caption{Results for the $mn$ model for $n=2$ in two different schemes. 
The index in column ``scheme", $4l$, $5l$, $6l$,  refers to the 
number of loops.  We report the coordinate of 
the FP $g_1$, $g_2$, the percentage of approximants that find the zero
($p_{\rm FP}$), and the percentage of approximants 
that indicate that the FP is stable ($p_{\rm st}$).
In the column ``info" we report the number of approximants that give
real (first number) and complex eigenvalues (second number).
}
\label{tabmn}
\end{table}

In order to check for the presence of additional FPs we considered the 
six-loop MZM expansions of Ref.~\cite{PV-00} and we generated 5-loop
$3d$-$\overline{\rm MS}$ expansions.  For the analysis we used 
the conformal-mapping method: the position of the Borel singularity
in the MZM scheme is reported in Ref.~\cite{PV-00}, while in the 
$3d$-$\overline{\rm MS}$ we used its trivial generalization. 
The two $\beta$ functions were
resummed by using several different approximants depending on two 
parameters, $b$ and $\alpha$ (see Ref.~\cite{CPV-00} for definitions).
For simplicity, each time we resummed the two $\beta$ functions 
by using the same $b$ and $\alpha$ and then determined their common zeroes. 
In principle, it would have been more natural to consider different values of 
$b$ and $\alpha$ for the two $\beta$ functions and all possible 
combinations. However, as we already tested in the 
$O(2)\otimes O(N)$ model, the two choices give fully equivalent results.
In the analysis we used $\alpha = -1,0,1,2$ and $b=4,6,\ldots,14$, 
which appeared 
to be an optimal choice. We report the results in Table \ref{tabmn}.
For comparison, we also performed the analysis for $m=2$, obtaining 
results completely consistent with those reported above. In the table 
we also give the percentage of cases in which a FP was found ($p_{\rm FP}$)
and in which this FP was stable $(p_{\rm st})$. Finally, we also 
indicate the number of cases in which the stability eigenvalues were real
or complex.

For $m=2$ and $m=3$ the presence of a new FP is unambiguous. Essentially
all considered approximants at five and six loops 
in both schemes give a stable FP. 
For $m=4$, the percentages are smaller, although the overall results are 
still in favor of a new stable FP. For $m\ge 5$ there is essentially 
no evidence. As far as the stability eigenvalues, for $m=2$ they
are complex, in agreement with the results for the 
$O(2)\otimes O(2)$ model \cite{PRV-01,CPS-02,CPPV-04}. For $m=3$ and 
$m=4$ the numerical results favor real eigenvalues instead.
It is interesting to note that the $3d$-$\overline{\rm MS}$ FPs lie at the 
boundary of the 
region in which the expansions are Borel summable,
$g_1 + g_2 > 0$. This is not the case for the MZM ones (Borel summability 
requires $R_{2m} g_1 + R_{m} g_2 > 0$). Thus, the 
$3d$-$\overline{\rm MS}$ results should be more reliable in these models. 

The $mn$ model for $n=2$ is invariant under the group 
$[SO(m)\oplus SO(m)]\otimes C_2$ where $C_2$ is the symmetry group of the 
square.
We now consider two quartic operators that break such a symmetry:
\begin{eqnarray}
P_1 & \equiv & (\Phi_1\cdot\Phi_1)^2 - (\Phi_2\cdot\Phi_2)^2, \\
P_2 & \equiv & (\Phi_1\cdot\Phi_2)^2 - 
     {1\over m} (\Phi_1\cdot\Phi_1) (\Phi_2\cdot\Phi_2), 
\end{eqnarray}
where the scalar products are taken in the $O(m)$ space. 
The first operator is the only quartic one that preserves the continuous 
symmetry and breaks $C_2 \to {\mathbb Z}_2\oplus {\mathbb Z}_2$, 
while the second preserves $C_2$ but breaks 
$SO(m)\oplus SO(m)\to SO(m)$. Note that in general $P_1$ mixes with
the lower-dimensional operator $\Phi_1\cdot\Phi_1 - \Phi_2\cdot\Phi_2$. 
Such a mixing should be taken into account in the MZM scheme, but does not 
occur in the massless $\overline{\rm MS}$ scheme. 
The operators $P_1$ and $P_2$  are relevant for the stability 
of the FPs of the $mn$ theory in larger models with smaller symmetry
group. 

We computed the anomalous dimensions of $P_1$ and $P_2$ at the 
new FPs of Table \ref{tabmn} by analyzing the corresponding 
5-loop $3d$-$\overline{\rm MS}$ series. The exponent $y_1$ was obtained 
from the analysis of the inverse
series $1/y_1$; the direct analysis of the series of $y_1$ was very
unstable. For $y_2$ we used instead the corresponding series. The 
results were not very stable and should be taken with caution. They are:

\begin{tabular}{ccc}
$m=2$: & $y_1 = 0.4(3)$, & $y_2 = -0.9(5)$; \\
$m=3$: & $y_1 = 0.2(2)$, & $y_2 = -1.0(8)$; \\
$m=4$: & $y_1 = 0.2(2)$, & $y_2 = -0.8(5)$. \\
\end{tabular}

It is interesting to note that the $mn$ model is a submodel of the 
multicritical Hamiltonian (\ref{HmcrNN}) for $r_1 = r_2$ and 
$f_{1,0} = f_{3,0}$ if we set $\Phi_{a1} = \phi_a$, $\Phi_{a2} = \psi_a$,
$f_{1,0} = f_{3,0} = g_{1,0} + g_{2,0}$, and $f_{2,0} = 2 g_{1,0}$. 
This implies that the new FPs may be relevant for the multicritical 
behavior of ${\cal H}_{\rm mcr,2}$. To investigate this possibility 
we must compute the anomalous dimension of the operator that breaks 
$[SO(m)\oplus SO(m)]\otimes C_2\to O(m)\oplus O(m)$, i.e., the 
operator $P_1$. As it can be seen, $y_1 > 0$ in all cases, 
indicating that the $mn$ FP is unstable in the full theory. Thus, the 
decoupled FP appears to be the only stable FP of the 
multicritical model (\ref{HmcrNN}) \cite{Aharony-02,CPV-03}.


\begin{thebibliography}{99}

\bibitem{ZJ-libro}
Zinn-Justin J.
Quantum Field Theory and Critical Phenomena, fourth edition. 
Clarendon Press, Oxford, 2001. 

\bibitem{PV-rev}
Pelissetto A.,  Vicari E. //
Phys. Rept., 2002,  vol. 368, No. 6, p.~549--727.

\bibitem{KNF-76}
Kosterlitz J.M., Nelson D.R., Fisher M.E. //
Phys. Rev. B, 1976, vol. 13, No. 1, p.~412--432.

\bibitem{CPV-04}
Calabrese P., Pelissetto A., Vicari E.
Multicritical behavior in frustrated spin systems with noncollinear order.
Preprint, cond-mat/0408130, 28 p.

\bibitem{Aharony-76}
Aharony A.
Dependence of universal critical behaviour on symmetry and range of 
interaction.
-- In: Phase Transitions and Critical Phenomena.
Vol. 6, edited by C. Domb and M.S. Green, New York, Academic, 1976, 
p. 357--424.

\bibitem{SD-89}
Schloms R., Dohm V. //
Nucl.\ Phys.\ B, 1989, vol. 328, No. 3, p.~639--663;
Phys. Rev. B, 1990, vol. 42, No. 10, p. 6142-6152; 
erratum Phys. Rev. B, 1992, vol. 46, No. 9, p. 5883.

\bibitem{KS-01} 
Kleinert H., Schulte-Frohlinde V. 
Critical Properties of $\phi^4$-Theories. 
Singapore, World Scientific, 2001.

\bibitem{BPV-03}
Butti A., Pelissetto A., Vicari E. // 
J. High Energy Phys., 2003, vol. 08, art. 029, p.~1--27.

\bibitem{Kawamura-88}
Kawamura H. //  Phys. Rev. B, 1988, vol. 38, No. 7, p. 4916--4928;
erratum   Phys. Rev. B, 1990, vol. 42, No. 4, p. 2610.

\bibitem{Kawamura-98}
Kawamura H. // J. Phys.: Condens. Matter, 1998,  vol. 10, No. 22, 
p.~4707--4754.

\bibitem{DMT-03}
Delamotte B., Mouhanna D., Tissier M. //
Phys. Rev. B, 2004, vol. 69, No. 13, art. 134413, p. 1--53.

\bibitem{CPPV-04}
Calabrese P., Parruccini P., Pelissetto A., Vicari E. 
Critical behavior of $O(2)\otimes O(N)$ symmetric models.
Preprint, cond-mat/0405667, 2004, 45 p.
(to appear in Phys. Rev. B).

\bibitem{PRV-01}
Pelissetto A., Rossi P., Vicari E. //
Phys. Rev. B, 2001,  vol. 63,  No. 14, art. 140414(R), p. 1--4.

\bibitem{CPS-02}
Calabrese P., Parruccini P., Sokolov A.I. //
Phys. Rev. B, 2002, vol. 66, No. 18, art. 180403(R), p. 1--4;
Phys. Rev. B, 2003, vol. 68, No. 9,  art. 094415, p. 1--8. 

\bibitem{ASV-95}
Antonenko S.A., Sokolov A.I., Varnashev K.B. //
Phys. Lett. A, 1995, vol. 208, No. 1-2, p.~161--164.

\bibitem{PRV-01b}
Pelissetto A., Rossi P., Vicari E. //
Nucl. Phys. B, 2001, vol. 607, No. 3, p.~605--634.

\bibitem{CP-03}
Calabrese P., Parruccini P. //
Nucl. Phys. B, 2004, vol. 679, No. 3, p.~568--596.

\bibitem{DPV-04}
De Prato M., Pelissetto A., Vicari E. 
The normal-to-planar superfluid transition in $^3$He.
Preprint, cond-mat/0312362, 2003, 19 p. 
(to appear in Phys. Rev. B, 2004, vol. 70).

\bibitem{TDM-00}
Tissier M., Delamotte B., Mouhanna D. //
Phys. Rev. Lett., 2000, vol. 84, No. 22, p.~5208--5211.

\bibitem{KW-01}
Kindermann M., Wetterich C. // 
Phys. Rev. Lett., 2001, vol. 86, No. 6, p.~1034-1037.

\bibitem{CPV-03}
Calabrese P., Pelissetto A., Vicari E. //
Phys. Rev. B, 2003, vol. 67, No. 5, art. 054505, p. 1--12.

\bibitem{Aharony-02}
Aharony A. // 
Phys. Rev. Lett., 2002, vol. 88, No. 5, art. 059703, p. 1.

\bibitem{CPV-00}
Carmona J.M., Pelissetto A.,  Vicari E. //
Phys. Rev. B, 2000, vol. 61, No. 22, p.~15136--15151.

\bibitem{DPV-03}
De Prato M., Pelissetto A., Vicari E. //
Phys. Rev. B, 2003, vol. 68, No. 09, art. 092403, p. 1--4.

\bibitem{CPV-02}
Calabrese P., Pelissetto A., Vicari E. //
Phys. Rev. E, 2002, vol. 65, No. 4, art. 046115, p.~1--16

\bibitem{CPV-O4nova}
Calabrese P., Pelissetto A., Vicari E.
The critical behavior of magnetic systems described by Landau-Ginzburg-Wilson 
field theories. --
In: Frontiers in Superconductivity Research, edited by Barry P. Martins. 
Hauppauge, NY, Nova Science, 2004.
Also as: Preprint, cond-mat/0306273, 2003, 29 p.

\bibitem{DHY-04}
Dudka M., Holovatch Yu., Yavors'kii T.
Universality classes of three-dimensional $mn$-vector models.
Preprint, cond-mat/0404217, 2004.

\bibitem{PV-00}
Pelissetto A., Vicari E. //
Phys. Rev.  B, 2000, vol. 62, No. 10, p. 6393--6409.



\end{thebibliography}
\end{document}